\documentclass[aps,prl,showpacs,onecolumn]{revtex4-1}
\usepackage[utf8]{inputenc}
\usepackage[english]{babel}
\usepackage{graphicx}
\usepackage{amsmath}
\usepackage{amsfonts}
\usepackage{amssymb}

\begin{document}

\title{Facets on the convex hull of $d$-dimensional Brownian and Lévy motion}
\date{\today}
\author{Julien Randon-Furling}
\email{Julien.Randon-Furling@univ-paris1.fr}
\affiliation{SAMM (EA 4543), Universit\'e Paris-1 Panth\'eon-Sorbonne, Centre Pierre Mend\`es-France, 90 rue de Tolbiac, 75013 Paris, France}
\author{Florian Wespi}
\email{Florian.Wespi@stat.unibe.ch}
\affiliation{IMSV, Universit\"at Bern, Sidlerstrasse 5, 3012 Bern, Switzerland}

\begin{abstract}
For stationary, homogeneous Markov processes (viz., L\'{e}vy processes, including Brownian motion) in dimension~$d\geq 3$, we establish an exact formula for the average number of $(d-1)$-dimensional facets that can be defined by $d$ points on the process's path. This formula defines a universality class in that it is independent of the increments' distribution, and it admits a closed form when $d=3$, a case which is of particular interest for applications in biophysics, chemistry and polymer science.

We also show that the asymptotical average number of facets behaves as $\langle \mathcal{F}_T^{(d)}\rangle \sim 2\left[\ln \left( T/\Delta t\right)\right]^{d-1}$, where $T$ is the total duration of the motion and $\Delta t$ is the minimum time lapse separating points that define a facet.
\end{abstract}


\maketitle

\section*{Introduction}

Stochastic processes' paths are common in the physical and natural sciences, in particular as models of actual motion~\cite{Einst,Chandra,Vis,OBen}, interfaces~\cite{havlin} or long proteins and other polymers~\cite{Edw,dGe,West,DoiEdw}. Spatial characteristics of stochastic processes thus come to reflect physically meaningful variables: home ranges of animals~\cite{Worton}, heights of surfaces~\cite{MC}, shapes of molecules~\cite{Haber,PRCDM}. 

One way to describe the spatial extent of a stochastic process's path is to examine its convex hull. That is, the smallest convex set enclosing all points visited by the process. In two dimensions, exact expressions for the values of geometric quantities such as the mean perimeter or the mean area of the convex hull have been found for Brownian motion \cite{Ta,El,BiLe,CHM} as well as for other types of stochastic motions~\cite{RMR,KLM,LGE}, for multi-walker systems~\cite{RFSMAC} and for confined configurations~\cite{chupeau,chupeau2}. In some cases, equivalent results exist in higher dimensional space~\cite{Kin2,RandonPhD,Davy,KLM,Eldan,MW}. 

Results~\cite{Rud,FouDes,Haber} pertaining to the shape of such random convex hulls appear scarcer though, despite their importance in biophysics: a protein's or an antibody's ability to play its part depends crucially on its shape~\cite{Cell,ashapes}. Convex hulls methods are central in the description of long-molecule configurations ~\cite{Holm,Meier,Lee,Stout,Wang2006}, leading to questions on the structure of these hulls, such as on their number of edges (in 2D) or facets (in 3D) and on the distribution of distances between the molecule and its convex hull's edges or facets~\cite{Badel,Coleman}.

Only recently, the average number of edges on the boundary of the convex hull of a vast class of planar random walks was computed. Remarkably, this was found to have a universal form: it grows logarithmically with time~\cite{Randon}, whatever the distribution of the infinitesimal increments along the process's path --~as long as these increments are independent and identically distributed (i.e. as long as the process is what is called a L\'evy process).

\begin{figure}
\includegraphics[scale=0.7]{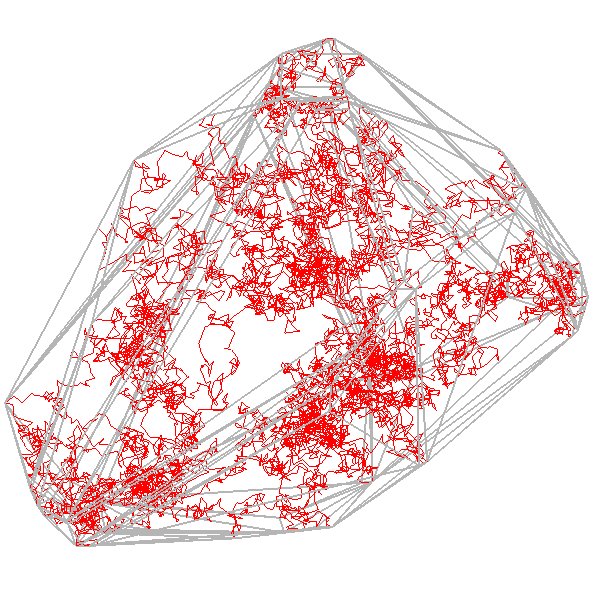}
 \caption{\label{3DHull} A sample realization of 3-dimensional Brownian motion and the facets on the boundary of its convex hull.}
 \end{figure}

In this paper, we show that the universality observed for the average number of edges in two dimensions extends to the average number of facets in $d\geq 3$~dimensions~(Fig.\ref{3DHull}). We establish an exact, integral formula for the average number of facets on the convex hull of a vast class of L\'evy processes in dimension~$d$. When $d=3$, this formula admits a closed form,
\begin{eqnarray*}
\langle \mathcal{F}_T^{(3)}\rangle = 2\left[\ln \left( T/\Delta t\right)\right]^2+4\left\lbrace\ln\left(T/\Delta t\right) \ln\left(1-\Delta t/T\right)\right.\\\left.-\mathrm{Li_2}\left(1-\Delta t/T\right)+\pi^2/12\right\rbrace,
\end{eqnarray*}
where $\mathrm{Li_2}$ is the dilogarithm function, $T$ is the total duration of the process and $\Delta t$ is a cut-off parameter representing an arbitrary minimal time separation between two extreme points on the process's path (see details in section~2). Asymptotics are obtained from the general $d$-dimensional formula,
\begin{equation}
\langle \mathcal{F}_T^{(d)}\rangle \sim 2\left[\ln \left( T/\Delta t\right)\right]^{d-1},
\end{equation}
which generalizes the two-dimensional case, where the average number of edges on the boundary of the convex hull was shown in~\cite{Randon} to be given by~$2\ln\left(T/\Delta t\right)$.

For the sake of clarity, we first examine the 3-dimensional discrete-time case. In section~2 we then establish an integral formula in the general $d$-dimensional, continuous-time case, before deriving asymptotics and obtaining a closed form when $d=3$ in secton~3. We illustrate our results with numerical simulations.

\section{Discrete-time case}
\label{sec:discrete-time-case}

In three-dimensional space, the convex hull of the successive positions of a random walker is almost surely made up of triangular facets. Following~\cite{Randon}, we shall compute the average number of such facets by computing the probability for a triangle defined by three points on the walker's path to form a facet of the convex hull~(see~Fig.~\ref{3DHull}).

Let $S_n$ denote the position of the walker after $n$ steps ($S_0=0$, $n\geq 3$). Given three integers $n_1 < n_2 < n_3$, the triangle joining $S_{n_1}$, $S_{n_2}$, and $S_{n_3}$ will be a facet of the walk's convex hull if the walker always stays on the same side of the plane defined by $S_{n_1}$, $S_{n_2}$, $S_{n_3}$. Let us choose a reference frame where the $xy$-plane coincides with the one defined by~$S_{n_1}$, $S_{n_2}$, $S_{n_3}$. Then the three-dimensional random walk always stays on the same side of the $xy$-plane if its $z$-coordinate is a one-dimensional random walk that: (i)~starts at some value --~say a positive one~-- and hits~$0$ for the first time at time $n_1$; (ii)~performs a positive excursion away from~$0$ between~$n_1$ and~$n_2$ (i.e. it hits~$0$ at these times but remains positive in between); (iii)~performs another positive excursion away from~$0$ between~$n_2$ and~$n_3$; and finally (iv)~leaves from~$0$ at time~$n_3$ and stays positive thereafter.

Relabeling $i=n_1$, setting $k_1=n_2-n_1$ and $k_2=n_3-n_2$, one may formally write the following formula for the mean number $\langle \mathcal{F}_N^{(3)}\rangle$ of facets \begin{equation}
  2\sum_{k_1=1}^{N-1} \sum_{k_2=1}^{N-k_1} \sum_{i=0}^{N-(k_1+k_2)} 
  m_{i}^{(0)} e_{i+k_1}^{(i)} e_{i+k_1+k_2}^{(i+k_1)} m_N^{(i+k_1+k_2)},
   \label{eq:GenFormDT}
  \end{equation}
with $m_{i}^{(0)}$, $e_{i+k_1}^{(i)}$, $e_{i+k_1+k_2}^{(i+k_1)}$, $m_N^{(i+k_1+k_2)}$ standing for the probabilities of each of the four parts~(i)-(iv) described above. The prefactor~$2$ accounts for the fact that there are two sides to every facet.

Since we are considering only random walks with independent and identically distributed jumps, we have $e_{i+k_1}^{(i)}=e^{(0)}_{k_1}\equiv e_{k_1}$ 
and $e_{i+k_1+k_2}^{(i+k_1)}=e^{(0)}_{k_2}\equiv e_{k_2}$ for all $i$, $k_1$, and $k_2$. Therefore, Eq.~(\ref{eq:GenFormDT}) becomes
\begin{equation} 
\label{eq:GenFormDT2}
 \langle \mathcal{F}_N^{(3)}\rangle =2\sum_{k_1=1}^{N-1} \sum_{k_2=1}^{N-k_1} e_{k_1} e_{k_2} 
  \sum_{i=0}^{N-(k_1+k_2)} m_{i}^{(0)} m_N^{(i+k_1+k_2)}.  
\end{equation}

Markovian independence allows one to stitch the parts of the $z$-random walk described above as pleases (because the cuts occur at special times, namely stopping times). In particular, stitching together the first and fourth parts produces a sub-walk of duration~$N-\left(k_1+k_2\right)$. This walk starts at some positive value and hits its minimum ($0$ by construction) at time~$i=n_1$. The probability associated with this sub-walk is just the probability for a random walk with~$N-\left(k_1+k_2\right)$ steps to hit its minimum at step~$i$ (see~\cite{Randon} for more details). Thus interpreting the product $m_{i}^{(0)}  m_N^{(i+k_1+k_2)}$ in Eq.~(\ref{eq:GenFormDT2}), one finds that
\begin{equation}
  \sum_{i=0}^{N-(k_1+k_2)}m_{i}^{(0)}  m_N^{(i+k_1+k_2)}=
 \sum_{i=0}^{N-(k_1+k_2)}\mathrm{Prob}\left(\mathrm{Min.\ is\ hit\ at\ }i\right)=1.\nonumber
\end{equation}
Substituting into Eq.~(\ref{eq:GenFormDT2}) yields
\begin{equation}
  \langle \mathcal{F}_N^{(3)}\rangle=2\sum_{k_1=1}^{N-1} \sum_{k_2=1}^{N-k_1} e_{k_1} e_{k_2}.
\end{equation}

There remains to compute the excursion probabilities $e_{k_1}$, $e_{k_2}$. As in~\cite{Randon}, let us notice that the excursion probability for a random walk is the probability that the walk, pinned at $0$ at times $0$ and $N$, visits only the positive half space. This is simply given by the probability that a bridge (that is, a random walk pinned at $0$ at times $0$ and $N$) attains its minimum at the initial step. The well-known uniform law for bridges (linked to the arcsine law for free random walks~\cite{Fitzsimmons,Kn,Kal}) therefore gives simply $e_{k}=1/k$. (Note that using generalizations of the arcsine and uniform laws~\cite{Barndorff}, one could treat the three-dimensional case here without resorting to a projection on one of the coordinates. We do it only for the sake of clarity.)

Finally, we obtain
\begin{equation}
\label{eq:3DDTRes}
  \langle \mathcal{F}_N^{(3)}\rangle=2\sum_{k_1=1}^{N-1} \sum_{k_2=1}^{N-k_1} \frac{1}{k_1 k_2}
  =2\sum_{\substack{k_1+ k_2 \leq N \\ k_1, k_2 \geq 1}}\frac{1}{k_1 k_2}.
\end{equation}

In higher dimension, the same reasoning is readily carried out. Counting facets that can be defined by $d$~points on the path of a random walk with $N\geq d$~steps, one obtains a $(d-1)$-fold sum
\begin{equation}
\label{eq:dDDTRes}
  \langle \mathcal{F}_N^{(d)}\rangle=2\sum_{\substack{k_1+ \dots + k_{d-1} \leq N \\ k_1,\dots ,k_{d-1} \geq 1}}
   \frac{1}{k_1\cdot k_2\cdots k_{d-1}}.
\end{equation}
While writing this paper, the result published recently in~\cite{Dimitri} came to our attention: this is similar to our~Eq.~(\ref{eq:dDDTRes}). Several other results pertaining to the discrete-time case are to be found in~\cite{Dimitri}, but none on the continuous-time case, which we now examine.

\section{Continuous-time case}
\label{sec:continuous-time-case}

Continuous-time analogs to stationary, homogeneous random walks are known as L\'evy processes~\cite{Bertoin}. They have independent, identically distributed (iid) increments and examples are Brownian motion, Cauchy processes, L\'evy-Smirnov processes and stable processes~\cite{LevyBrownian}.

In two dimensions, the mean number of edges on the convex hull of such processes' paths exhibits a universal form, $\langle \mathcal{F}_T^{(2)}\rangle =2\ln \left(T/\Delta t\right)$, where $T$ is the total duration of the process. The cutoff parameter $\Delta t$ may be interpreted as the minimal time separation, between two points on the process's path, for them to be examined as potential endpoints of an edge on the convex hull~(see~\cite{BMCHEdges,Randon} for a detailed explanation).

Given the universality observed in both the discrete-time case and the two-dimensional continuous time case, one expects a simple continuous version of~Eq.(\ref{eq:dDDTRes}) to hold in the $d$-dimensional continuous-time case.

This may be proven using the same strategy as in the previous section. In dimension~$d$, a facet will be a $(d-1)$-dimensional object lying in a hyperplane and defined by $d$ points on the process's path. One thus splits the path into $d+1$~independent parts, and then computes the probability density (pdf) associated with each part before multiplying them together and integrating. When $d=3$, the $4$~parts and their associated pdfs are:
\begin{itemize}
  \item from time $0$ up to time~$\tau$, the path lies on one side only (say side~$+$) of a (hyper)plane $\mathcal{H}$, and it hits $\mathcal{H}$ at time $\tau$; write $p_{\tau}^{(0)}$ for the corresponding pdf;
  \item from time $\tau$ up to time $\tau + k_1$, the path is an excursion away from $\mathcal{H}$ (in side~$+$) and is pinned on~$\mathcal{H}$ at times~$\tau$ and~$\tau + k_1$; write $f(\tau,\tau+k_1)$ for the pdf; 
  \item from time $\tau + k_1$ up to time $\tau + k_1 + k_2$, the path is an excursion away from~$\mathcal{H}$ (in side~$+$) and is pinned on~$\mathcal{H}$ at times $\tau + k_1$ and $\tau + k_1 +k_2$; write $f(\tau + k_1,\tau + k_1+k_2)$ for the pdf;
  \item from time~$\tau +k_1+k_2$ up to time~$T$, the path lies in side~$+$ of~$\mathcal{H}$ given that it is pinned on $\mathcal{H}$ at time~$\tau +k_1+k_2$; write~$p_T^{(\tau +k_1+k_2)}$ for the pdf.
\end{itemize} 

Formally, one obtains the following integral
\begin{equation}\label{eq:3DGenEq}
 \langle \mathcal{F}_T^{(3)}\rangle=2\int_{\Delta t}^{T-\Delta t}\int_{\Delta t}^{T-k_1} \int_{0}^{T-(k_1+k_2)}\ dk_1\ dk_2\ d\tau
  p_{\tau}^{(0)}f(\tau,\tau+k_1)f(\tau +k_1,\tau +k_1+k_2) p_T^{(\tau+k_1+k_2)}.
\end{equation}

Since L\'{e}vy processes have stationary and independent increments, we have, just as in the discrete-time case, for all~$\tau$, $k_1$ and $k_2$: $f(\tau,\tau+k_1)=f(0,k_1)\equiv f(k_1)$ and $f(\tau+k_1,\tau+k_1+k_2)=f(0,k_2)\equiv f(k_2)$. Also, the product $p_{\tau}^{(0)} p_T^{(\tau+k_1+k_2)}$ can be interpreted as giving 
the pdf of the time $\tau$ at which a similar process of duration $T-(k_{1}+k_{2})$ attains its minimum. Thus, 
\begin{equation}
  \int_{\tau=0}^{T-(k_1+k_2)}p_{\tau}^{(0)} p_T^{(\tau+k_1+k_2)}d\tau=1,
\end{equation} 
and the combined integral in Eq.~(\ref{eq:3DGenEq}) reduces to
\begin{equation}
  \langle \mathcal{F}_T^{(3)}\rangle =2\int_{\Delta t}^{T-\Delta t}
  \int_{\Delta t}^{T-k_1}dk_1\ dk_2 f(k_1) f(k_2).  
\end{equation}

Lastly, $f(k)$ may be viewed as the pdf that the sojourn time of a process on a given side of a (hyper)plane to which it is pinned (at both endpoints) is equal to the full duration $k$. This means, $f(k)$ is the pdf of the sojourn time of a L\'evy bridge, which is known to be uniform in higher dimensions, just as it is in dimension~$1$~\cite{Fitzsimmons}. Therefore $f(k)=1/k$, and we obtain
\begin{equation}
\label{eq:3dCTInt}
  \langle \mathcal{F}_T^{(3)}\rangle = 2\int_{\substack{k_1+ k_2 \leq T \\ k_1, k_2 \geq \Delta t}}\frac{dk_1\,dk_2}{k_1 k_2}.
\end{equation}

In higher dimension, Eq.~(\ref{eq:3DGenEq}) becomes a $d$-fold integral, accounting for facets defined by $d$ points and characterised by $d-1$ excursions:
\begin{equation}
  \langle \mathcal{F}_T^{(d)}\rangle = 
  2 \int_{\Delta t}^{T-(d-2)\Delta t}
  \int_{\Delta t}^{T-(d-3)\Delta t-k_1} \cdots \int_{\Delta t}^{T-(k_1+\dots + k_{d-2})}
  \frac{dk_1\cdots  dk_{d-2}\,dk_{d-1} }{k_1 k_2\cdots k_{d-1}}, \nonumber
\end{equation}
that is,
\begin{equation}
\label{eq:dDGenInt}
  \langle \mathcal{F}_T^{(d)}\rangle  =2 \int_{\substack{k_1+ \dots + k_{d-1} \leq T \\ k_1,\dots ,k_{d-1} \geq \Delta t}}
  \frac{dk_1\cdots  dk_{d-2}\,dk_{d-1} }{k_1 k_2\cdots k_{d-1}}.
\end{equation}
In the three-dimensional case, it is possible to obtain a closed form for this integral formula, as detailed in the next section.

\section{Closed formula and asymptotics}
Starting from Eq.~(\ref{eq:3dCTInt}), we change variables to $y_1=k_1$, $y_2=k_1+k_2$. We find
\begin{equation}
\label{eq:1st2DInt}
\langle \mathcal{F}_T^{(3)}\rangle=2\int_{\Delta t}^{T-\Delta t}\,\frac{dy_1}{y_1}\, \int_{y_1+\Delta t}^{T}\, \frac{dy_2}{y_2-y_1},
\end{equation}
in which the second integral is easily computed. This leads to:
\begin{equation}
\label{eq:2nd2DInt}
\langle \mathcal{F}_T^{(3)}\rangle=2\int_{\Delta t}^{T-\Delta t} \frac{dy_1}{y_1} \left[\ln \left(\frac{T}{\Delta t}\right)+\ln\left(1-\frac{y_1}{T}\right)\right].
\end{equation}
Setting $z=y_1/T$ allows one to compute the remaining integral (see also~\cite{Maximon,Zagier}) and yields our main result,
\begin{equation}
\label{eq:MR}
\langle \mathcal{F}_T^{(3)}\rangle = 2\left[\ln \left( T/\Delta t\right)\right]^2+4\left\lbrace\ln\left(T/\Delta t\right) \ln\left(1-\Delta t/T\right)-\mathrm{Li_2}\left(1-\Delta t/T\right)+\pi^2/12\right\rbrace,\nonumber
\end{equation}
where $\mathrm{Li_2}$ is the dilogarithm function. As $\mathrm{Li_2\left(1^-\right)}$ is finite, the first term on the right hand side of~Eq.(\ref{eq:MR}) is also clearly the leading term when $T/\Delta t$ becomes large. Thus,
\begin{equation}
\label{eq:asMR}
\langle \mathcal{F}_T^{(3)}\rangle \sim 2\left[\ln \left( T/\Delta t\right)\right]^2.
\end{equation}
This asymptotical behaviour may readily be directly extracted from~Eq.~(\ref{eq:2nd2DInt}), focusing on the $\ln \left(T/\Delta t\right)$ term in the integrand. This transposes straightforwardly, by simple iteration, to the $d$-dimensional integral in~Eq~(\ref{eq:dDGenInt}), leading to our second main result,
\begin{equation}
\langle \mathcal{F}_T^{(d)}\rangle \sim 2\left[\ln \left( T/\Delta t\right)\right]^{d-1}.
\end{equation}
This generalizes in a very simple manner the two-dimensional result~\cite{Randon} and the discrete-time result~\cite{Dimitri}.

\begin{figure}
\includegraphics[scale=0.33]{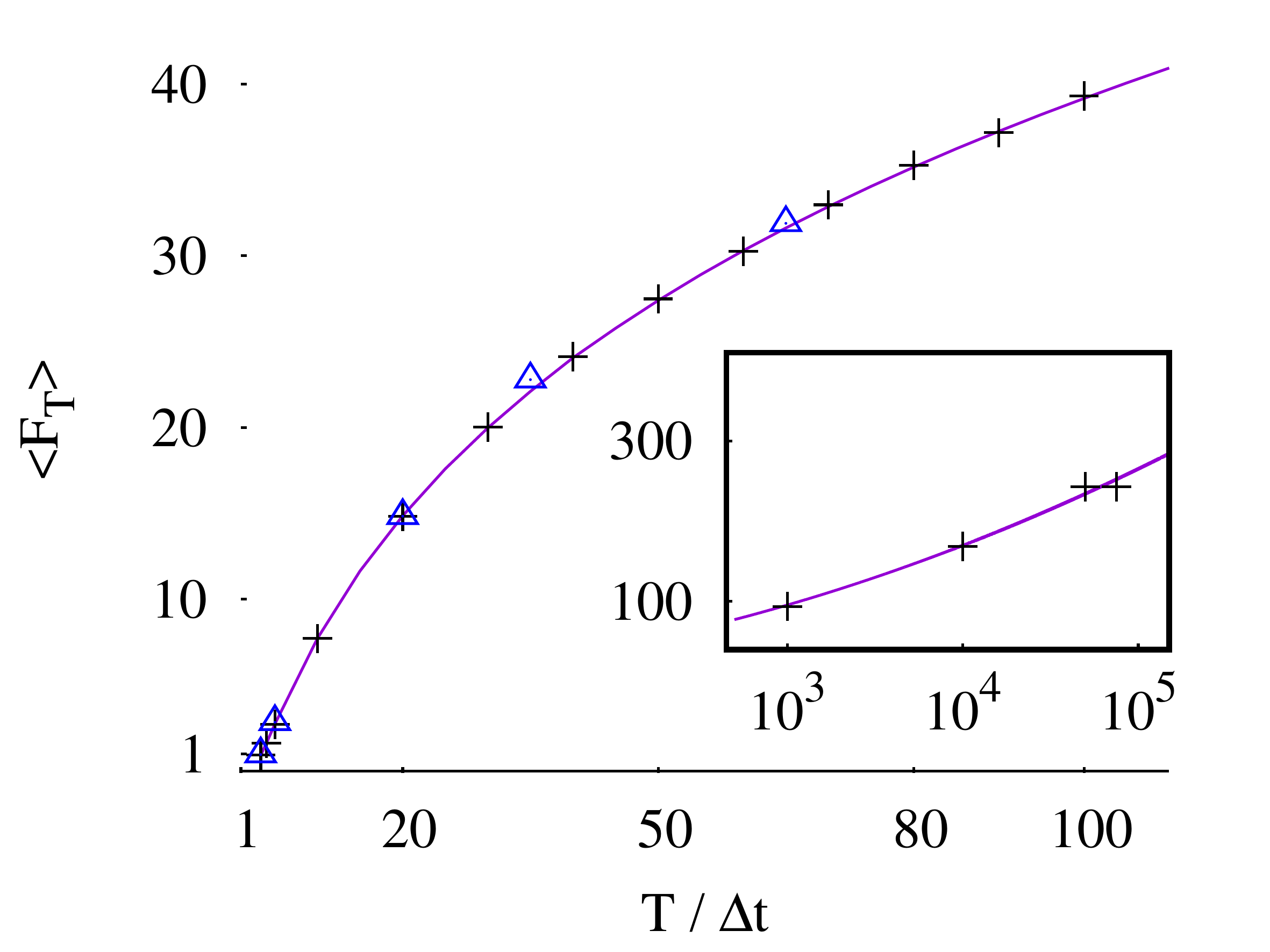}
 \caption{\label{NumSim} Numerical simulations for the average number of triangular facets on the convex hulls of some $3$-dimensional L\'evy processes (with $10^4$ and $10^3$ sample paths’ realizations): standard Brownian motion (pluses), and a stable process with stability index $1.5$ (triangles). The solid line corresponds to the exact formula~Eq.~(\ref{eq:MR}). The inset shows the asymptotics for large values of $T/\Delta t$, from~Eq~(\ref{eq:asMR}), with numerical simulations of 3-dimensional Brownian motion.}
 \end{figure}

Both the exact formula (Eq.~(\ref{eq:MR})) and the asymptotical behaviour (Eq.~(\ref{eq:asMR})) are confirmed in numerical simulations, as can be seen on~Fig.~\ref{NumSim}. It is also the case in dimension $d\geq 4$.

\section{Conclusion}

The exact formula established here for the average number of $(d-1)$-dimensional facets on the convex hull of a general $d$-dimensional L\'evy process shows how universal certain aspects of the shape of spatial random processes are. Indeed this formula is valid not only for $d$-dimensional Brownian motion but for a vast class of L\'evy processes, including all stable processes. This is reminiscent of the universality observed in the Sparre Andersen theorem~\cite{SA1,SA2}, to which the formula established here is linked via the arcsine and the uniform laws. To be more specific about the class of processes for which our formula holds: one needs the process to be truly $d$-dimensional and to have infinitely many increments on any time interval. This may not be the case for instance with compound Poisson processes. However, in many applications, one is dealing with Brownian motion or stable (self-similar) processes, which are covered here. Further work is needed to explore the cases not included, it may well be that they fall into the same universality class.

We will also seek to adapt the method introduced here to address multi-processes systems: what is the average number of facets on the global convex hull of $N$ independent spatial Brownian motions? One could also expect a large universality class in this context, and it will be of particular interest to see how the number of walkers intervenes on the $2\left[\ln \left(T/\Delta t\right)\right]^{d-1}$ asymptotics in the long time limit.

\bibliography{3dBib.bib}

\end{document}